\documentclass[twocolumn]{aastex701}
\usepackage{amsmath}
\usepackage{booktabs}
\usepackage{stix2}

\newcommand{\sgr}{SGR~J1935}

\begin{document}

\title{Joint Observation of SGR J1935+2154 with \textit{Insight}-HXMT and KM40m during the active episode of October 2022}

\correspondingauthor{Xiao-Bo Li, Long-Fei Hao, Shao-Lin Xiong}
\email{lixb@ihep.ac.cn, haolongfei@ynao.ac.cn, xiongsl@ihep.ac.cn}

\author[0000-0001-8664-5085]{Wang-Chen Xue}
\affiliation{State Key Laboratory of Particle Astrophysics, Institute of High Energy Physics, Chinese Academy of Sciences, Beijing 100049, China}
\affiliation{University of Chinese Academy of Sciences, Chinese Academy of Sciences, Beijing 100049, China}
\email{xuewc@ihep.ac.cn}

\author[0009-0006-5506-5970]{Wen-Jun Tan}
\affiliation{State Key Laboratory of Particle Astrophysics, Institute of High Energy Physics, Chinese Academy of Sciences, Beijing 100049, China}
\affiliation{University of Chinese Academy of Sciences, Chinese Academy of Sciences, Beijing 100049, China}
\email{tanwj@ihep.ac.cn}

\author[0009-0000-8142-8612]{Yu-Xiang Huang}
\affil{Yunnan Observatories, Chinese Academy of Sciences, Kunming 650216, China}
\affil{University of Chinese Academy of Sciences, Chinese Academy of Sciences, Beijing 100049, China}
\affil{Key Laboratory for the Structure and Evolution of Celestial Objects, Chinese Academy of Sciences, Kunming 650216, China}
\email{huangyuxiang@ynao.ac.cn}

\author[0000-0003-4585-589X]{Xiao-Bo Li\textsuperscript{*}}
\affiliation{State Key Laboratory of Particle Astrophysics, Institute of High Energy Physics, Chinese Academy of Sciences, Beijing 100049, China}
\email{lixb@ihep.ac.cn}

\author{Long-Fei Hao\textsuperscript{*}}
\affil{Yunnan Observatories, Chinese Academy of Sciences, Kunming 650216, China}
\affil{Key Laboratory for the Structure and Evolution of Celestial Objects, Chinese Academy of Sciences, Kunming 650216, China}
\email{haolongfei@ynao.ac.cn}

\author[0000-0002-4771-7653]{Shao-Lin Xiong\textsuperscript{*}}
\affiliation{State Key Laboratory of Particle Astrophysics, Institute of High Energy Physics, Chinese Academy of Sciences, Beijing 100049, China}
\email{xiongsl@ihep.ac.cn}

\author[0000-0002-6540-2372]{Ce Cai}
\affiliation{College of Physics and Hebei Key Laboratory of Photophysics Research and Application, Hebei Normal University, Shijiazhuang, Hebei 050024, China}
\email{caice@hebtu.edu.cn}

\author[0009-0008-8053-2985]{Chen-Wei Wang}
\affil{State Key Laboratory of Particle Astrophysics, Institute of High Energy Physics, Chinese Academy of Sciences, Beijing 100049, China}
\affil{University of Chinese Academy of Sciences, Chinese Academy of Sciences, Beijing 100049, China}
\email{cwwang@ihep.ac.cn}

\author[0009-0008-5068-3504]{Yue Wang}
\affil{State Key Laboratory of Particle Astrophysics, Institute of High Energy Physics, Chinese Academy of Sciences, Beijing 100049, China}
\affil{University of Chinese Academy of Sciences, Chinese Academy of Sciences, Beijing 100049, China}
\email{yuewang@ihep.ac.cn}

\author[0000-0002-1435-0883]{Ke-Jia Lee}
\affil{Department of Astronomy, Peking University, Beijing 100871, China}
\affil{National Astronomical Observatories, Chinese Academy of Sciences, Beijing 100101, China}
\affil{Yunnan Observatories, Chinese Academy of Sciences, Kunming 650216, China}
\affil{Beijing Laser Acceleration Innovation Center, Huairou, Beijing, 101400, China}
\email{kjlee@pku.edu.cn}

\author[0000-0002-5031-8098]{Heng Xu}
\affil{National Astronomical Observatories, Chinese Academy of Sciences, Beijing 100101, China}
\email{hengxu@bao.ac.cn}

\author[0000-0002-8097-3616]{Peng Zhang}
\affil{State Key Laboratory of Particle Astrophysics, Institute of High Energy Physics, Chinese Academy of Sciences, Beijing 100049, China}
\affil{College of Electronic and Information Engineering, Tongji University, Shanghai 201804, China}
\email{zhangp97@ihep.ac.cn}

\author{Ming-Yu Ge}
\affil{State Key Laboratory of Particle Astrophysics, Institute of High Energy Physics, Chinese Academy of Sciences, Beijing 100049, China}
\email{gemy@ihep.ac.cn}

\author{Hao-Xuan Guo}
\affil{State Key Laboratory of Particle Astrophysics, Institute of High Energy Physics, Chinese Academy of Sciences, Beijing 100049, China}
\affil{Department of Nuclear Science and Technology, School of Energy and Power Engineering, Xi'an Jiaotong University, Xi'an 710049, China}
\email{guohx@ihep.ac.cn}

\author{Yue Huang}
\affil{State Key Laboratory of Particle Astrophysics, Institute of High Energy Physics, Chinese Academy of Sciences, Beijing 100049, China}
\email{huangyue@ihep.ac.cn}

\author{Cheng-Kui Li}
\affil{State Key Laboratory of Particle Astrophysics, Institute of High Energy Physics, Chinese Academy of Sciences, Beijing 100049, China}
\email{lick@ihep.ac.cn}

\author[0009-0004-1887-4686]{Jia-Cong Liu}
\affil{State Key Laboratory of Particle Astrophysics, Institute of High Energy Physics, Chinese Academy of Sciences, Beijing 100049, China}
\affil{University of Chinese Academy of Sciences, Chinese Academy of Sciences, Beijing 100049, China}
\email{liujiacong@ihep.ac.cn}

\author{Yang-Zhao Ren}
\affil{State Key Laboratory of Particle Astrophysics, Institute of High Energy Physics, Chinese Academy of Sciences, Beijing 100049, China}
\affil{School of Physical Science and Technology, Southwest Jiaotong University, Chengdu 611756, China}
\email{renyz@ihep.ac.cn}

\author[0000-0003-2957-2806]{Shuo Xiao}
\affil{School of Physics and Electronic Science, Guizhou Normal University, Guiyang 550001, China}
\affil{Guizhou Provincial Key Laboratory of Radio Astronomy and Data Processing, Guizhou Normal University, Guiyang 550001, China}
\email{xiaoshuo@gznu.edu.cn}

\author[0000-0001-9217-7070]{Sheng-Lun Xie}
\affil{Institute of Astrophysics, Central China Normal University, Wuhan 430079, China}
\affil{State Key Laboratory of Particle Astrophysics, Institute of High Energy Physics, Chinese Academy of Sciences, Beijing 100049, China}
\email{xiesl@ihep.ac.cn}

\author{Shu-Xu Yi}
\affil{State Key Laboratory of Particle Astrophysics, Institute of High Energy Physics, Chinese Academy of Sciences, Beijing 100049, China}
\email{sxyi@ihep.ac.cn}

\author[0009-0002-6411-8422]{Zheng-Hang Yu}
\affil{State Key Laboratory of Particle Astrophysics, Institute of High Energy Physics, Chinese Academy of Sciences, Beijing 100049, China}
\affil{University of Chinese Academy of Sciences, Chinese Academy of Sciences, Beijing 100049, China}
\email{zhyu@ihep.ac.cn}

\author[0009-0007-6192-0213]{Jin-Peng Zhang}
\affil{State Key Laboratory of Particle Astrophysics, Institute of High Energy Physics, Chinese Academy of Sciences, Beijing 100049, China}
\affil{University of Chinese Academy of Sciences, Chinese Academy of Sciences, Beijing 100049, China}
\email{zhangjinpeng@ihep.ac.cn}

\author[0000-0001-5348-7033]{Yan-Qiu Zhang}
\affil{School of Physics and Electronic Science, Guizhou Normal University, Guiyang 550001, China}
\affil{State Key Laboratory of Particle Astrophysics, Institute of High Energy Physics, Chinese Academy of Sciences, Beijing 100049, China}
\affil{University of Chinese Academy of Sciences, Chinese Academy of Sciences, Beijing 100049, China}
\email{yanqiuzhang@gznu.edu.cn}

\author[0009-0001-7226-2355]{Chao Zheng}
\affil{State Key Laboratory of Particle Astrophysics, Institute of High Energy Physics, Chinese Academy of Sciences, Beijing 100049, China}
\affil{TIANFU Cosmic Ray Research Center, Chengdu, Sichuan, China}
\email{zhengchao97@ihep.ac.cn}

\author{Shi-Jie Zheng}
\affil{State Key Laboratory of Particle Astrophysics, Institute of High Energy Physics, Chinese Academy of Sciences, Beijing 100049, China}
\email{zhengsj@ihep.ac.cn}

\author{Shu-Mei Jia}
\affil{State Key Laboratory of Particle Astrophysics, Institute of High Energy Physics, Chinese Academy of Sciences, Beijing 100049, China}
\email{jiasm@ihep.ac.cn}

\author{Xiang Ma}
\affil{State Key Laboratory of Particle Astrophysics, Institute of High Energy Physics, Chinese Academy of Sciences, Beijing 100049, China}
\email{max@ihep.ac.cn}

\author{Jin Wang}
\affil{State Key Laboratory of Particle Astrophysics, Institute of High Energy Physics, Chinese Academy of Sciences, Beijing 100049, China}
\email{jinwang@ihep.ac.cn}

\author{Hai-Sheng Zhao}
\affil{State Key Laboratory of Particle Astrophysics, Institute of High Energy Physics, Chinese Academy of Sciences, Beijing 100049, China}
\email{zhaohs@ihep.ac.cn}

\author[0000-0001-9834-2196]{Yong Chen}
\affiliation{State Key Laboratory of Particle Astrophysics, Institute of High Energy Physics, Chinese Academy of Sciences, Beijing 100049, China}
\email{ychen@ihep.ac.cn}

\author[]{Cong-Zhan Liu}
\affiliation{State Key Laboratory of Particle Astrophysics, Institute of High Energy Physics, Chinese Academy of Sciences, Beijing 100049, China}
\email{liucz@ihep.ac.cn}

\author{Yu-Peng Xu}
\affil{State Key Laboratory of Particle Astrophysics, Institute of High Energy Physics, Chinese Academy of Sciences, Beijing 100049, China}
\email{xuyp@ihep.ac.cn}

\author[0000-0003-0274-3396]{Li-Ming Song}
\affil{State Key Laboratory of Particle Astrophysics, Institute of High Energy Physics, Chinese Academy of Sciences, Beijing 100049, China}
\affil{University of Chinese Academy of Sciences, Chinese Academy of Sciences, Beijing 100049, China}
\email{songlm@ihep.ac.cn}

\author[0000-0001-5586-1017]{Shuang-Nan Zhang}
\affil{State Key Laboratory of Particle Astrophysics, Institute of High Energy Physics, Chinese Academy of Sciences, Beijing 100049, China}
\affil{University of Chinese Academy of Sciences, Chinese Academy of Sciences, Beijing 100049, China}
\email{zhangsn@ihep.ac.cn}

\begin{abstract}
SGR J1935+2154 is the unique magnetar so far from which fast radio bursts have been detected.
In October 2022, it resumed its burst activity, and we implemented a dedicated target-of-opportunity (ToO) observation on it from Oct. 13th to Nov. 1st, 2022 (about 940 ks in total) with \textit{Insight}-HXMT, while the KM40m radio telescope observed this source for about 1400 hours since Oct. 15th.
We searched the LE, ME, and HE data of \textit{Insight}-HXMT in the overlapping observation time windows with the KM40m radio telescope and revealed 60 magnetar X-ray bursts (MXBs), while KM40m only detected 1 radio burst.
In particular, we find that there is an X-ray burst on October 21 (denoted as MXB 221021) temporally associated with this radio burst. Interestingly, this association event shows very different morphology from those X-ray and radio association events from this source reported before (e.g., MXB/FRB 200428).
Moreover, we systematically analyzed the temporal and spectral properties of the sample of MXBs during this observation and found that %
the (radio-associated) MXB 221021 shows some different properties from other MXBs without associated radio bursts.
These findings shed new light on the physical mechanisms of X-ray bursts and radio burst emission in magnetars.
\end{abstract}


\section{Introduction} \label{sec:intro}
Magnetars are a special subclass of neutron stars, characterized by their extraordinarily powerful magnetic fields that exceed ${\sim}10^{14}\,$Gauss \citep{Duncan92Magnetar, Thompson95SGR, Thompson96SGR}.
The hallmarks of magnetars are their X/$\gamma$-ray transient activities, including (short) bursts, giant flares, and outbursts \citep{Rea11Magnetar}.
At present, about 30 magnetars have been identified \citep{Olausen14MagnetarCatalog}, which can be categorized into two classes: those with high burst rates, known as soft gamma-ray repeaters (SGRs), and those with fewer burst occurrences, known as anomalous X-ray pulsars (AXPs) \citep{Kaspi17Magnetar}.
The persistent emissions and burst activities of magnetars are believed to be powered by the decay and instability of their intense magnetic field \citep[e.g.][]{Mereghetti15Magnetar, Turolla15Magnetar, Kaspi17Magnetar, Esposito21Magnetar}.

SGR~J1935+2154 (hereafter \sgr{}) has been one of the most active magnetars over the past decade. It was discovered in 2014, following a short magnetar X-ray burst (MXB) that triggered \textit{Swift}/BAT \citep{Stamatikos14GCN}.
Follow-up observations with \textit{Swift}/XRT, \textit{Chandra}, and XMM-\textit{Newton} confirmed its magnetar nature, revealing its spin period of ${\sim}$3.24$\,$s, spin-down rate of ${\sim}1.43\,{\times}\,10^{-11}\,$s/s, and surface dipolar magnetic field strength of ${\sim}2.2\,{\times}\,10^{14}\,$Gauss \citep{Israel16SGRJ1935}.
\sgr{} is a Galactic magnetar, its distance measurements are within 1.5--12.5$\,$kpc \citep{kothes_radio_2018, zhong_distance_2020, zhou_revisiting_2020, bailes_multifrequency_2021}.
\sgr{} has exhibited multiple active phases, producing bursts and outbursts in 2015, 2016, 2019, 2020, and 2022 \citep{Kozlova16IntermediateFlare, Younes17Outbursts, Borghese20SGRJ1935, Lin20SGRJ1935, Lin20BurstProperties, Kaneko21GBM1935Forest, Yang21GBM1935Burst, Cai22BurstCatalog, Cai22SpecCatalog, Xie22SGRJ1935Periodicity, Ibrahim24SGRJ1935, Shao24SGRJ1935Outburst, Xie2025ApJS, Fu2025SGRJ1935NICER}.

Fast radio bursts (FRBs) are mysterious sources in radio astronomy \citep{Lorimer07FRB, Thornton13FRB, Xiao21FRB, Petroff22FRB, Zhang23FRB}. A seminal discovery on the origins of FRBs was made during the 2020 outburst of \sgr{}.
A bright Galactic FRB (denoted as FRB 200428) \citep{CHIME20FRB200428, STARE220FRB200428} was found to temporally coincide with a magnetar X-ray burst (MXB 200428) from \sgr{} \citep{HXMT21FRB200428, INTEGRAL20FRB200428,  Konus21FRB200428, AGILE21FRB200428}, and the two narrow peaks of MXB are well aligned with the radio pulse of this FRB \citep{HXMT21FRB200428, INTEGRAL20FRB200428}.
Detailed analysis of the relation between narrow X-ray pulses and radio pulses further proves the consistency of the X-ray emission and radio emission \citep{0428peak}.
In particular, a quasi-periodic oscillation (QPO) signal at $\sim 40$~Hz was discovered in this burst \citep{0428QPO}.
Thus, this event provided compelling evidence for the idea that magnetars can produce FRBs (e.g., \citealt{Popov10sgrfrb, Katz16sgrfrb, Wadiasingh20sgrfrb, Metzger17sgrfrb}).

Despite this groundbreaking discovery, the regions of emission and the mechanisms behind the radio bursts produced by magnetars remain subjects of ongoing debate (e.g., \citealt{Lin20NoPulsedRadio, Younes21BroadbandXray, Zhu23RadioPulsarPhase, Zhang16frb, Lu20frb, Katz14frb, Usov00frb, Lyubarsky14frb, Waxman17frb}), primarily due to the sparsity of the observational data.
During the active phases of \sgr{}, MXBs' temporal and statistical properties have been comprehensively studied \citep{Xiao1935lag, Xiao1935soc, Xiao1935pds, Xiao1935pulseSOC, Xiao1935evolution,2025ApJ...985..211X}.
Continued multi-wavelength monitoring of \sgr{} and other magnetars is crucial to elucidate the emission mechanisms connecting X-ray and radio bursts, to determine whether the radio bursts are generated by magnetospheric processes or external shocks, and to explore the potential diversity in burst energetics and morphologies across different magnetar outbursts (e.g., \citealt{Margalit20FRBs, Beloborodov23Theory, Wadiasingh23MagnetarRadioModel, Xie2024ApJ}).

\sgr{} re-entered a new outburst phase in October 2022, prompting a series of multi-wavelength follow-up observations (e.g. \citealt{Atel15745, GCN32698, ATel15667, ATel15681, ATel15686, ATel15697, ATel15698, ATel15707}).
In particular, an increasing number of MXBs were found to be associated with radio bursts, including MXB 221014, the first GECAM detection of an FRB-associated MXB, which also marked the second case of MXB-FRB association \citep{ATel15682,Wang2026MXB221014}, and MXB 221120, which was identified as a peculiar FRB-associated MXB detected by GECAM \citep{Tan2026MXB221120}. Furthermore, the \textit{Insight}-HXMT mission \citep{ATel15708} detected a weak MXB (denoted as MXB 221021 in this work) associated with a radio burst discovered by the Kunming 40-meter radio telescope (KM40m) at Yunnan Observatories \citep{Atel1021km40}. Moreover, a sub-threshold MXB candidate was reported by the \textit{Fermi}/GBM \citep{ATel1201GBM} in relation to a radio burst identified by CHIME \citep{ATel1201CHIME}.
Although these radio bursts are not as intense as FRB 200428, these observations still enable a more comprehensive investigation into the association between the X-ray and radio bursts, which could shed light on the underlying physical processes involved in these events.

In this paper, we report on the simultaneous X-ray and radio observations of \sgr{}'s outburst in October 2022 with \textit{Insight}-HXMT \citep{Zhang20HXMTOverview} and the Kunming 40-meter radio telescope (KM40m; \citealt{KM40m}). With a focus on the X-ray data, we searched for MXBs in the HXMT ToO observation data, analyzed the properties of MXBs, and emphasized the case study of the MXB associated with the radio counterpart on Oct 21, 2022 (i.e., MXB 221021).

This paper is organized as follows. Section \ref{sec:obs-and-data} gives an overview of the observations. Section \ref{sec:data-analysis} discusses the data selection and methods employed for the analysis of X-ray and radio bursts. Our findings and their implications for the understanding of magnetar dynamics and FRB sources are discussed in Section \ref{sec:res-and-discuss}. We summarize our work in Section \ref{sec:summary}.

\section{Observations and Data Reduction}
\label{sec:obs-and-data}

\begin{figure*}[htb]
    \centering
    \includegraphics[width=1.0\textwidth]{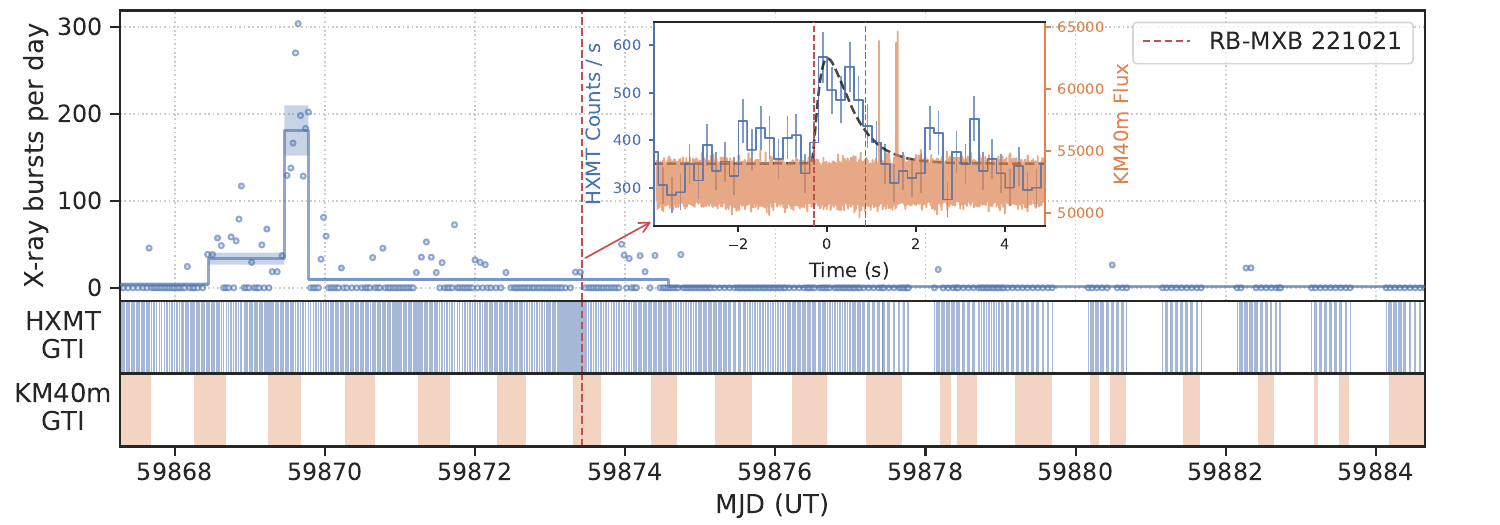}
    \caption{
        The joint observation history of \sgr{} by HXMT and KM40m. The red dashed line indicates the occurrence of the MXB 221021 associated with RB 221021.
        For the temporal comparison, the HXMT and KM40m event times shown here have been converted to geocentric arrival times (i.e., corrected for the light-travel-time difference between the observatory and the Earth center), and the KM40m radio arrival times have been de-dispersed to infinite frequency using the best-fit DM.
        \textbf{Top}: the data points are the observed burst rates during the good time intervals (GTIs) of HXMT observations, and the step line is the burst rate estimated from the Bayesian blocks algorithm (see text). The inset panel shows the light curves of the MXB observed by HXMT and the RB observed by KM40m, respectively. The black dashed curve shows the best-fit FRED function to the HXMT data.
        \textbf{Middle}: the HXMT GTIs are marked by the shaded regions in blue.
        \textbf{Bottom}: the periods during which KM40m was monitoring \sgr{} are denoted by the shaded areas in orange.
    }
    \label{fig:burst-history}
\end{figure*}

During the outburst phase of \sgr{} in October 2022, HXMT conducted Target of Opportunity (ToO) observations to \sgr{} in the X-ray band, starting from 2022-10-13T04:51:38 to 2022-11-01T15:48:26 (UTC), with an effective exposure of ${\sim}$940$\,$ks.
HXMT mainly consists of three collimated telescopes, the low energy X-ray telescope covering 1--10$\,$keV (LE; \citealt{Chen20LE}), the medium energy X-ray covering 10--35$\,$keV (ME; \citealt{Cao20ME}), and the high energy X-ray telescope covering 28--250$\,$keV (HE; \citealt{Liu20HE}). KM40m performed dedicated observations to \sgr{} in the radio band centered at a frequency of 2.245$\,$GHz with a bandwidth of 110$\,$MHz, from 2022-10-15T05:53:22 to 2023-03-22T01:07:54 (UTC), with a total exposure of ${\sim}$1400$\,$h.
The overlapping observation periods of the two telescopes are shown in Figure \ref{fig:burst-history}.
On Oct. 21, 2022, at 10:01:45 UTC, HXMT detected MXB 221021 \citep{ATel15708}, which was found to be temporally associated with a radio burst observed by KM40m (denoted as RB 221021, \citealt{ATel15707}).

Applying the same burst search and verification methods as in \cite{Cai22BurstCatalog} to HXMT observation data, over 300 MXB candidates from \sgr{} were identified \citep{Lu24TemporalSpectral}.
We further refined our search algorithm and identified 516 MXBs (W.-C. Xue et al., in prep.), 60 of which occurred during periods of overlapping X-ray and radio observations, including MXB 221021.
The burst history is shown in Figure \ref{fig:burst-history}, and the information on the 60 bursts is listed in Table \ref{tab:burst-analysis}.

The data of KM40m were recorded with a ROACH2 based system, and the software package \texttt{BEAR} \citep{Men19BEAR} was used for radio burst signal searching.
The detection significance of this radio burst, RB~221021, is approximately 20-$\sigma$. RB~221021 is composed of multiple narrow pulses distributed within $\sim0.5$~s. We identify three prominent sub-bursts (RB~20221021A--C), whose measured properties are summarized in Table~\ref{tab:radio}. We note that RB~221021 is the only radio burst detected by KM40m during the overlapping observation period with HXMT.

\begin{deluxetable}{lccccc}
\tablecaption{Measured properties of the three prominent sub-bursts of RB~221021 detected by KM40m.\label{tab:radio}}
\tablehead{
\colhead{Event} & \colhead{Instrument} & \colhead{TOA} & \colhead{DM} & \colhead{Fluence} & \colhead{Width} \\
\colhead{} & \colhead{} & \colhead{(MJD)} & \colhead{(pc cm$^{-3}$)} & \colhead{(Jy ms)} & \colhead{(ms)}
}
\startdata
RB~20221021A & KM40m & 59873.4179010232 & $332.3^{+0.3}_{-0.3}$ & $37^{+20}_{-7}$ & $1.11^{+0.14}_{-0.14}$ \\
RB~20221021B & KM40m & 59873.4179054777 & $332.9^{+0.4}_{-0.4}$ & $11^{+6}_{-2}$ & $0.35^{+0.06}_{-0.06}$ \\
RB~20221021C & KM40m & 59873.4179059012 & $332.9^{+0.4}_{-0.4}$ & $13^{+7}_{-2}$ & $0.35^{+0.06}_{-0.06}$ \\
\enddata
\end{deluxetable}

\section{X-ray Data Analysis} \label{sec:data-analysis}
\subsection{Temporal Analysis} \label{section:temporal-analysis}

\begin{figure}[htb]
    \centering
    \includegraphics[width=0.49\textwidth]{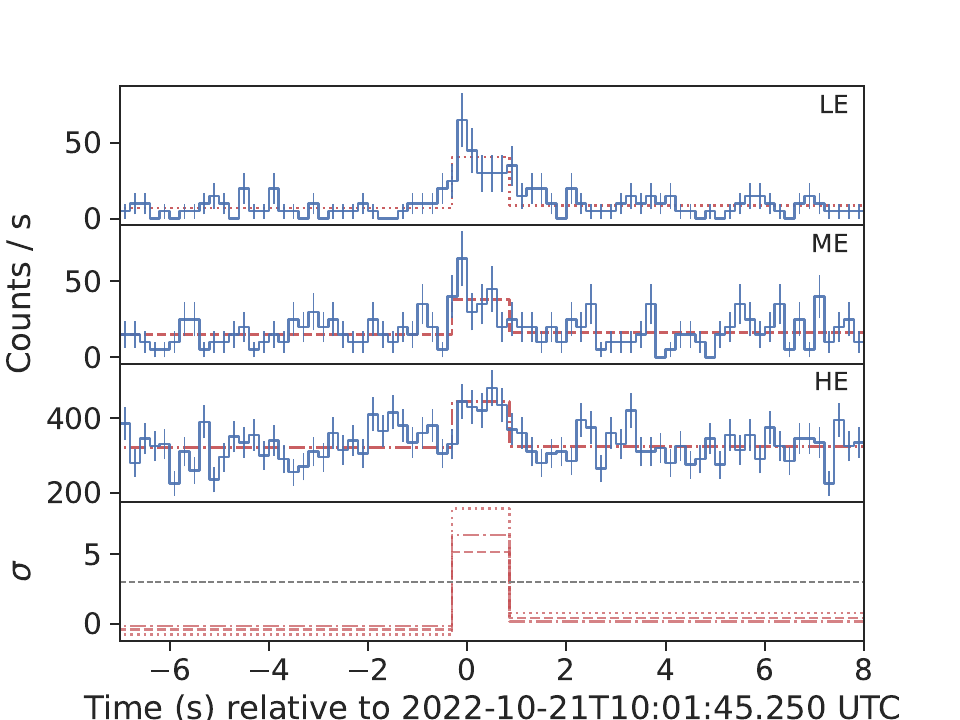}
    \caption{
        Light curves and Bayesian block representations of MXB 221021.
        The top three panels show the light curves (blue) and the corresponding Bayesian blocks (red) from three telescopes.
        The fourth panel presents the Bayesian block significance for LE (dotted), ME (dashed), and HE (solid), evaluated against the background. The gray dashed line marks the 3-$\sigma$ threshold.
    }
    \label{fig:bayesian-blocks}
\end{figure}

We estimate the duration of the X-ray burst based on the change points in the count rate identified by the Bayesian block algorithm \citep{Scargle13BayesianBlocks}.
The event time sequences of LE, ME, and HE are jointly fed to the Bayesian blocks algorithm to identify the global change points of the count rate.
The overall false positive rate (FPR) of change points is controlled by iteratively adjusting the prior parameter of the algorithm until the overall FPR converges to a 3-$\sigma$ level (<2.7\textperthousand).
The identified change points divide the light curve into a series of blocks.
The longest block is chosen as the initial background block for background estimation.
The next longest block is then tested for its significance against the background by the Li-Ma significance \citep{Li83Significance}.
If the significance is below the 3-$\sigma$ threshold, then this block is classified as a background block, and the background is updated with this block included.
If the significance exceeds the 3-$\sigma$ threshold, two cases are considered: If the duration between the block and the trigger time is within 3.24$\,$s, then it is identified as a burst block; otherwise, it is classified as a block that originates from background variation and is excluded from further analysis.
This process is applied iteratively to the remaining blocks, yielding an exact separation between the burst and the background intervals.
For example, Figure \ref{fig:bayesian-blocks} illustrates the segmentation of the light curve of MXB 221021.
The duration of the burst $\Delta T$ is therefore determined by the difference of two change points that separate the burst blocks from the background blocks.

To assess the coincident rate of MXB 221021 and RB 221021, we first estimate the X-ray burst rate.
Assuming that the number of detected bursts within a specific time interval follows a Poisson distribution, we can estimate the burst rate using Bayesian blocks, taking into consideration the gaps between GTIs.
The estimation of the burst rate is shown in top panel of Figure \ref{fig:burst-history}.
The estimated X-ray burst rate during the occurrence of RB 221021 is 10 bursts per day, which yields a chance probability of $1.8\times10^{-4}$ (equivalent to 3.6-$\sigma$ significance) for one or more X-ray bursts occurring in coincidence with the observed RB 221021 within a single magnetar spin period.
Therefore, the low probability of chance provides statistical evidence in favor of a genuine temporal association between MXB 221021 and RB 221021, suggesting a plausible physical connection that warrants further investigation.

To further quantify the relative timing between MXB~221021 and RB~221021, we modeled the HXMT X-ray light curve of MXB~221021 with a fast-rise exponential-decay (FRED) profile. The fit provides an estimate of the X-ray peak time $t_{\rm peak}^{X}=T_0 + 5.6_{-167.9}^{+191.9} \times 10^{-3}$ s. For the radio burst, the peak epoch $t_{\rm peak}^{R}$ was measured from the dedispersed KM40m time series. After referring both HXMT and KM40m times to the geocentric arrival time, and dedispersing the radio arrival times to infinite frequency, we compute the peak-to-peak offset as $\Delta t_{\rm peak}\equiv t_{\rm peak}^{R}-t_{\rm peak}^{X}$, with an uncertainty $\sigma_{\Delta}=\sqrt{\sigma_{X}^{2}+\sigma_{R}^{2}}\simeq \sigma_{X}$. This procedure yields a radio delay of $\Delta t_{\rm peak}=1.15_{-0.17}^{+0.19}$ s.

\subsection{Spectral Analysis}

For each X-ray burst detected by HXMT, we extracted time-integrated source spectra and the corresponding background spectra from the identified burst intervals and their associated background intervals, as described in Section~\ref{section:temporal-analysis}. Data from all three telescopes of HXMT are used, including the NaI detector of HE (denoted as HE/NaI), ME, and LE data. Together they cover a wide energy range from 2--250 keV, which is beneficial to constrain the MXB spectrum. For MXB 221021, we additionally extracted the CsI data of HE (denoted as HE/CsI) spectral data \citep{Zheng2025CsI} to better constrain the spectral shape at higher energies.

We fit the data with several commonly used empirical models, including:
\begin{itemize}
    \item The power-law function (PL),
    \begin{equation}
        N_\mathrm{PL}(E) \propto E^{-\alpha},
    \end{equation}
    where $\alpha$ is the photon index.

    \item The blackbody function (BB),
    \begin{equation}
        N_\mathrm{BB}(E) \propto \frac{E^2}{\exp(E/kT)-1},
    \end{equation}
    where $kT$ is the temperature of the blackbody.

    \item The optically-thin thermal bremsstrahlung model (OTTB),
    \begin{equation}
        N_\mathrm{OTTB}(E) \propto E^{-1} \exp\left(-\frac{E}{kT_\mathrm{e}}\right),
    \end{equation}
    where $kT_\mathrm{e}$ is the temperature of thermal electrons.

    \item The power-law function with a high-energy exponential cutoff (CPL),
    \begin{equation}
        N_\mathrm{CPL}(E) \propto E^{-\alpha} \exp\left(-\frac{E}{E_\mathrm{c}}\right),
    \end{equation}
    where $\alpha$ is the photon index, and $E_\mathrm{c}$ is the $e$-folding energy associated with exponential cutoff. Note that when $\alpha$ < 2, the peak of the $vF_v$ spectrum of CPL occurs at $E_\mathrm{p}=(2-\alpha)\ E_\mathrm{c}$.

    \item The combination of a non-thermal model and a blackbody, i.e., OTTB/PL/CPL+BB.
    \item The combination of double blackbodies BB+BB.
\end{itemize}

The above models are multiplied with a photoelectric absorption model with Wisconsin cross-sections (WABS; \citealt{Morrison83WAbs}) to account for the absorption by Galactic interstellar medium.
The equivalent hydrogen column density $\eta_\mathrm{H}$ of WABS for most spectra is not well constrained because of the limited photon counts in the low-energy band.
Therefore, we fix $\eta_\mathrm{H}$ to $2.79 \times 10^{22}\,\mathrm{atoms}/\mathrm{cm}^2$ during the fit, which is the same value adopted in previous studies \citep{HXMT21FRB200428, Cai22SpecCatalog}.

To account for the Poisson fluctuation in both the source and background spectra, we apply $W$-statistics for fitting these spectra \citep{Wachter79Wstat, Arnaud96XSPEC}.
Estimation with $W$-statistics can be biased when the background information is insufficient \citep{Buchner24SpecStats}.
A rule of thumb is to adjust the spectral channel binning such that the background spectra contain a minimum of five counts per channel bin.
To ensure the accuracy of our fitting results, we binned up the spectra using the optimal binning method \citep{Kaastra16OptimalBinning} along with the requirement of each bin containing a minimum of five background counts.

After the fit, we assess the validity of the fitted models using the parametric bootstrap method \citep{Efron94Bootstrap}.
We generate a set of $10^4$ synthetic spectra based on the fitted model and refit each with the same model.
The resulting distribution of $W$-statistics from these fits enables us to calculate the bootstrap $p$-value for the model applied to the observed burst spectra.
A model is considered to provide an adequate fit to the data if its bootstrap $p$-value exceeds 0.05.

To determine the most appropriate model while accounting for model complexity, we compare the relative goodness-of-fit of different models by employing the Akaike information criterion with sample size correction (AICc; \citealt{Akaike74AIC, Sugiura78AICc}).
When comparing a simple model with a complex model, we adopt the complex model only if $\Delta\mathrm{AICc} \equiv \mathrm{AICc}_{\rm simple}-\mathrm{AICc}_{\rm complex} > 2$; otherwise, we select the simpler model.

\begin{figure}[htbp]
    \centering
    \includegraphics[width=0.49\textwidth]{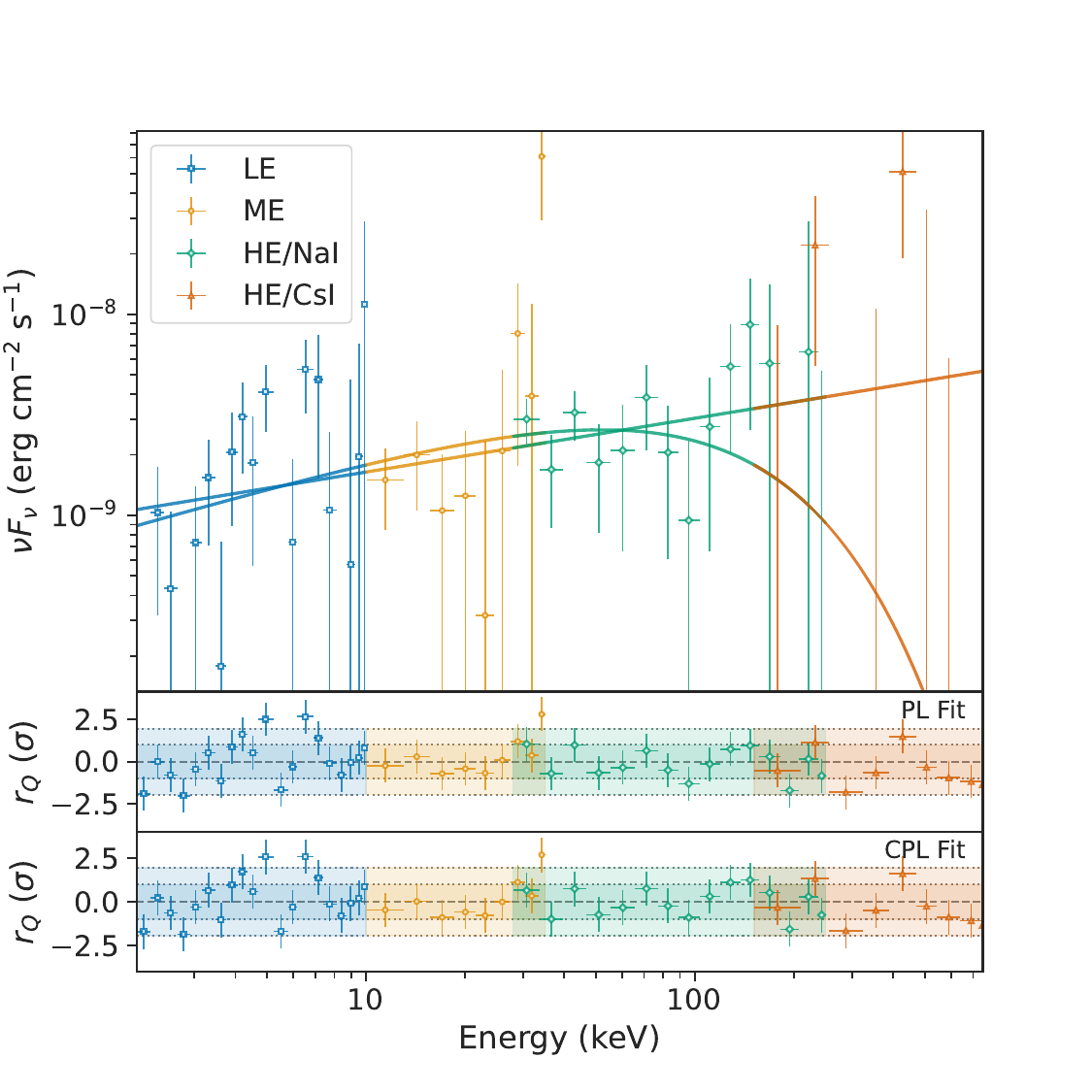}
    \caption{Spectral fits of RB–MXB~221021 with the PL and CPL models. The top panel shows the best-fit PL and CPL spectra, together with the $\nu F_\nu$ data points unfolded using the PL model. The last two panels present the residuals of the PL and CPL fits to the folded spectral data.}
    \label{fig:spec}
\end{figure}

\begin{figure*}[http]
\centering
\includegraphics[width=0.8\textwidth]{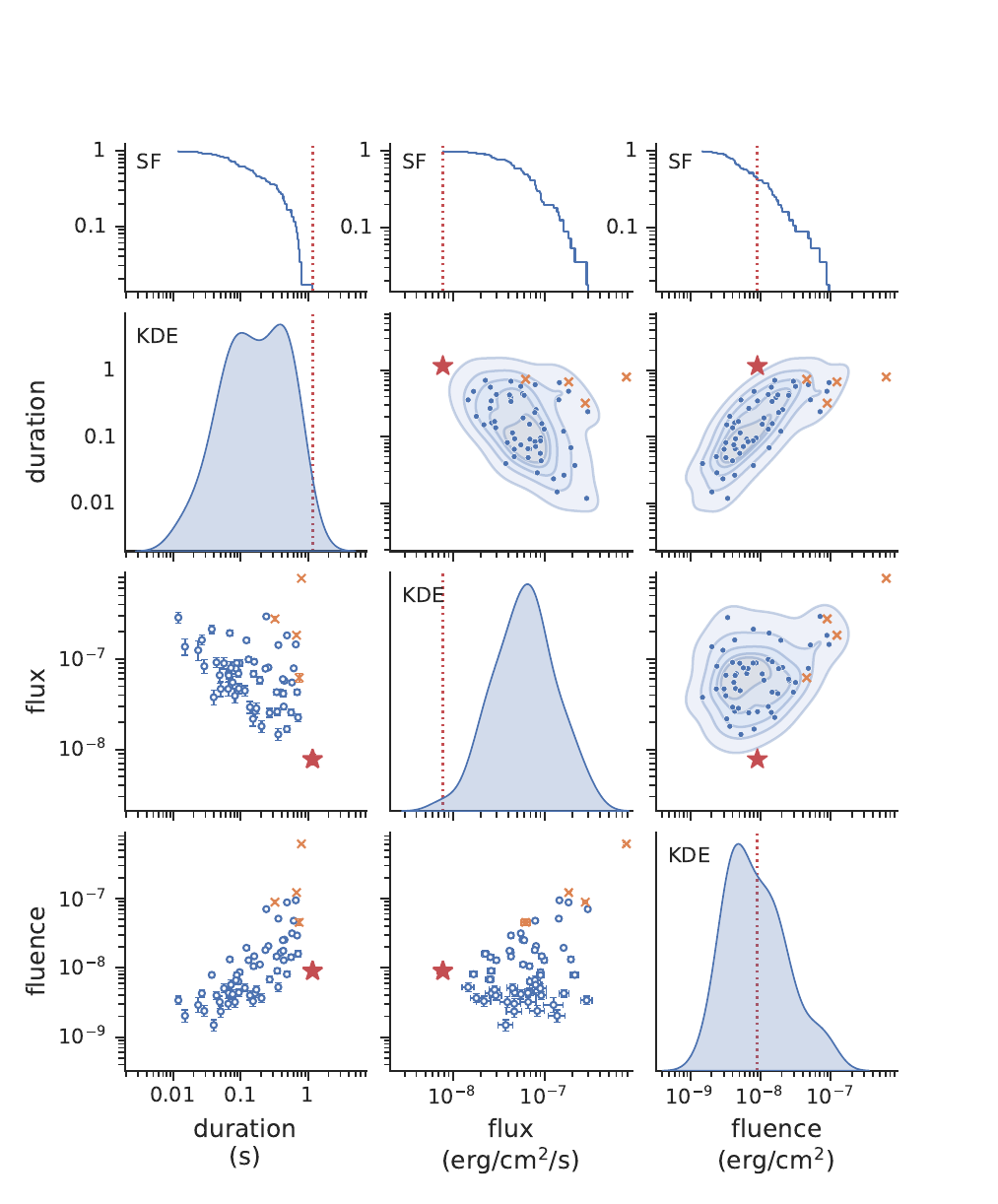}
\caption{The distribution and correlation of duration, fluence, and flux of 60 bursts from the selected model. The red star and red dashed line represent the RB-MXB 221021. The crosses in each panel represent the unreliable results due to the saturation of the HE telescope.}
\label{fig:burst-correlation}
\end{figure*}

\begin{figure*}[htbp]
    \centering
    \includegraphics[width=1\textwidth]{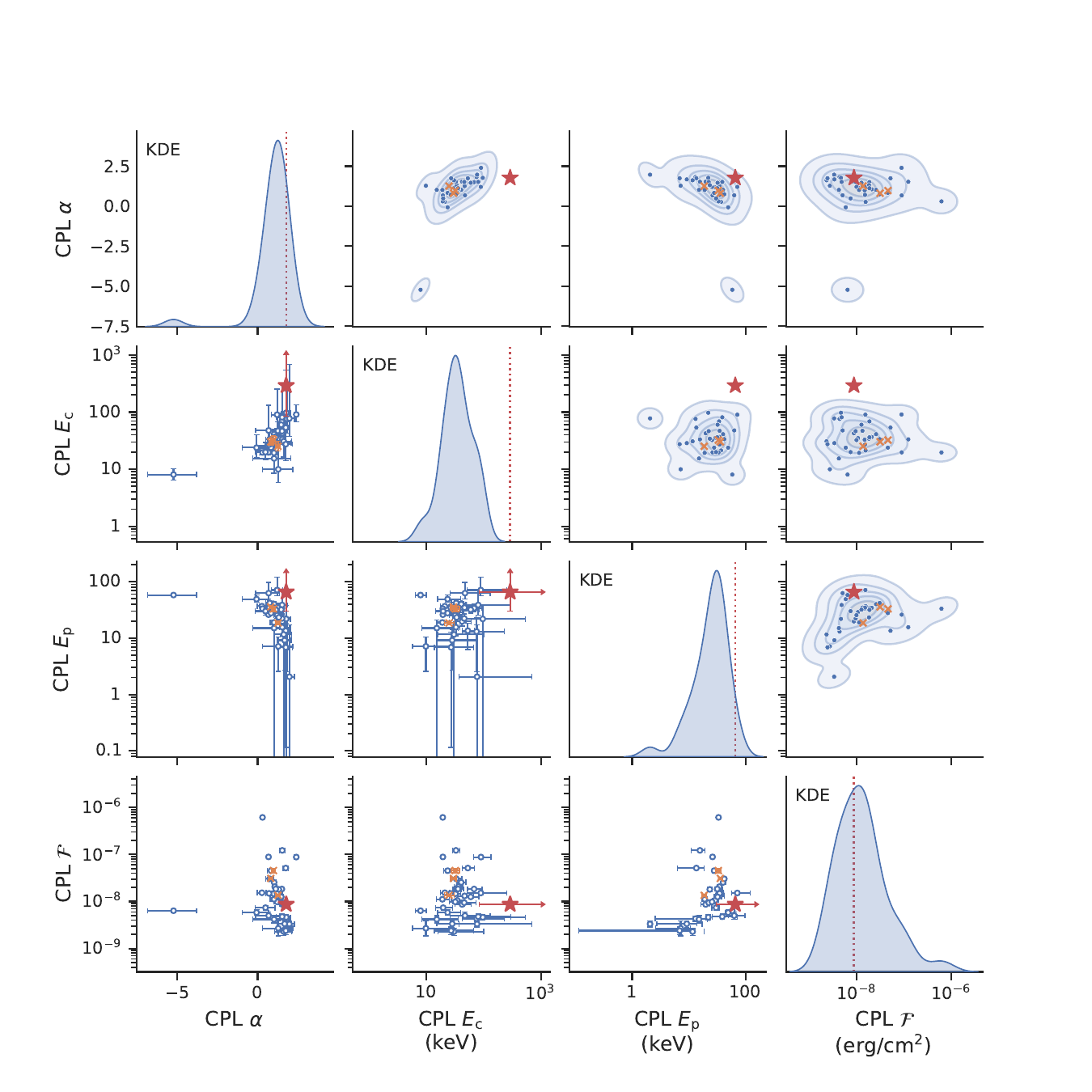}
    \caption{The distribution and correlation between CPL spectral parameters and burst properties. The red star and red dashed line represent the RB-MXB 221021. The crosses in each panel represent the unreliable results due to the saturation of the HE telescope.}
    \label{fig:correlation}
\end{figure*}

\section{Results and Discussion} \label{sec:res-and-discuss}

The temporal and spectral fitting results for the 60 MXB bursts detected during the contemporaneous observations by HXMT and KM40m are summarized in Table~\ref{tab:burst-analysis}.
For each burst, we report the best-fit parameters for the OTTB, PL, and CPL models, the BB and BB+BB models, as well as the hybrid OTTB/PL/CPL+BB model.
Following the procedure described above, we first assess model adequacy using the bootstrap $p$-value and require $p_{\rm boot}>0.05$.
Among the statistically adequate candidates, we select the preferred model by minimizing AICc.
The selected model is indicated in the ``Selected Model'' column of Table~\ref{tab:burst-analysis}.

Among the 60 bursts analyzed, 14 are best described by a PL model, 7 by a CPL model, 18 by an OTTB model, 1 by a single BB model, 6 by a BB+BB model, and 14 by a hybrid (OTTB/PL/CPL)+BB model.
Overall, non-thermal--dominated bursts (PL or CPL) account for $\sim35\%$ of the sample, implying that a thermal component is present in the majority of bursts in our dataset, broadly consistent with previous studies \citep{GBM_catalog_2022_Oct}.
This diversity in preferred spectral forms suggests that the burst emission may arise from multiple physical channels and/or emission sites within the magnetosphere, and that their relative contributions can vary substantially from burst to burst.

\subsection{The radio-X-ray associated event: RB-MXB~221021}

Prior to this work, two radio--high-energy association events from \sgr{} have been reported: FRB/MXB~200428 \citep{2020ATel13696....1Z, 2020ATel13686....1T, CHIME20FRB200428, STARE220FRB200428, HXMT21FRB200428, INTEGRAL20FRB200428, Konus21FRB200428, AGILE21FRB200428} and FRB/MXB~221014 associated with FRB~20221014A \citep{ATel15681, ATel15682, ATel15686, ATel15697}. For FRB/MXB~200428, the radio and high-energy peaks are nearly simultaneous after correcting for dispersion, with reported offsets at the millisecond level ($\sim$3--6~ms; see \citealt{CHIME20FRB200428, HXMT21FRB200428, INTEGRAL20FRB200428, Konus21FRB200428,0428peak}). For FRB/MXB~221014, the radio and high-energy peaks were also reported to be nearly simultaneous at the millisecond level after correcting for dispersion \citep{ATel15682, ATel15697,Wang2026MXB221014}.

In this work, we identify one X-ray burst, MXB~221021, that is temporally associated with the radio burst RB~221021 detected by KM40m.
The radio burst occurs $\sim$1~s after the peak time of the X-ray burst after correcting for dispersion, corresponding to $\sim 1/3$ of the spin period of \sgr{}.
While the exact delay depends on the adopted timing reference (e.g., X-ray peak vs.\ onset; radio peak vs.\ centroid) and on the uncertainty budget (including barycentering and radio dedispersion), the observed phase-scale offset suggests that the radio and X-ray emission are not strictly co-temporal in this event.
Such a $\sim$second-scale lag is orders of magnitude larger than the millisecond-scale offsets reported for FRB/MXB~200428 and FRB/MXB~221014 and can be challenging to reconcile with scenarios predicting near-simultaneity between the coherent radio burst and the prompt X-ray spike, unless additional geometric and/or propagation effects are considered \citep{Lu2020FRB, Israel2021}.

RB~221021 consists of three prominent sub-bursts, with fluences of $11$--$37$~Jy~ms and durations of $\sim0.35$--$1.11$~ms (Table~\ref{tab:radio}).
For comparison, FRB~200428 was orders of magnitude brighter in fluence: CHIME/FRB (400--800~MHz) reported fluences of $480$ and $220$~kJy~ms with widths of $0.585$~ms and $0.335$~ms, respectively \citep{CHIME20FRB200428}, while STARE2 (1400~MHz) measured a fluence of $1.5$~MJy~ms and a width of $0.61$~ms \citep{STARE220FRB200428}.
At lower fluence levels, several \sgr{} radio bursts without detected contemporaneous high-energy counterparts have been reported, including a burst detected by FAST (1250~MHz) on 2020 April 30 with a fluence of $60$~mJy~ms and a width of $1.966$~ms \citep{2020ATel13699....1Z}, and two bursts on 2020 May 24 detected with the 25-m RT1 telescope (1260--1388~MHz) at Westerbork with fluences of $24$ and $112$~Jy~ms and widths of $0.87$ and $0.96$~ms \citep{2021NatAs...5..414K}.
CHIME/FRB subsequently detected three additional bursts on 2020 October 8, spanning fluences of $\sim6$--$900$~Jy~ms and widths of $\sim0.2$--$1$~ms \citep{Giri2023SGRJ1935CHIME}. CHIME/FRB also detected two bright bursts in 2022: FRB~20221014A with a fluence of 9.7~kJy~ms and a width of 1.48~ms, and FRB~20221201A with a fluence of 23.7~kJy~ms and a width of 2.02~ms \citep{Giri2023SGRJ1935CHIME}.
Because fluence and width measurements are not strictly standardized across instruments and analyses, these comparisons are necessarily approximate. Nevertheless, RB~221021 has ms-scale widths comparable to previously reported \sgr{} radio bursts, and its fluence is far below FRB~200428-like events while remaining within the broad range of non-FRB-like bursts.

We performed time-integrated spectral analysis for MXB~221021, and the fitting results are shown in Figure~\ref{fig:spec}.
The spectrum is best described by a PL model with photon index $\alpha=1.86\pm0.09$, indicating a non-thermal and relatively soft X-ray spectrum.
When fitted with a CPL model, only the best-fit and lower bound of $E_{\rm c}$ and $E_{\rm p}$ can be constrained within the HXMT energy range, suggesting that the cutoff/peak energy is likely beyond (or only marginally sampled by) our bandpass.

From the PL fit, the X-ray fluence of MXB~221021 in 8--200~keV is $\mathcal{F}_{\rm X} = 9\times10^{-9}\ {\rm erg\ cm^{-2}}$.
To compare the energetics between radio and X-rays with consistent dimensions, we compute the radio energy fluence $\mathcal{F}_{\rm R}$ (in ${\rm erg\ cm^{-2}}$) from the measured radio fluence $\Phi_{\rm R}$ (in ${\rm Jy\ ms}$) via
\begin{equation}
\mathcal{F}_{\rm R} \simeq \Phi_{\rm R}\times 10^{-23}\ \frac{{\rm erg}}{{\rm s\ cm^2\ Hz}}\times 10^{-3}\ {\rm s}\times \Delta\nu,
\end{equation}
where $\Delta\nu$ is the effective bandwidth in Hz.
Using the reported radio fluence of RB~221021 and the KM40m bandwidth ($\Delta\nu\simeq110$~MHz), we obtain $\mathcal{F}_{\rm R}/\mathcal{F}_{\rm X}\sim10^{-9}$.
This ratio is below the typical range ($\sim10^{-5}$--$10^{-8}$) reported for other magnetar radio/X-ray events \citep{chen2020frb}, suggesting an unusually low radio-to-X-ray energy partition for RB-MXB~221021 under the adopted definitions.
We note that $\mathcal{F}_{\rm R}$ depends on the effective bandwidth and the (generally uncertain) radio spectral shape across the band; nevertheless, the conclusion that the radio-to-X-ray ratio is comparatively small is robust at the order-of-magnitude level.

\subsection{Magnetar X-ray burst population correlations}

Figure~\ref{fig:burst-correlation} shows the distributions and correlations among burst duration, fluence, and flux for the HXMT MXB samples.
We observe a positive association between burst duration and fluence, consistent with previous studies of magnetar bursts \citep{Gogus01SGR1806, Gavriil04AXP1E, Lin20BurstProperties}.
Compared with the Fermi/GBM sample \citep{GBM_catalog_2022_Oct}, our sample extends to substantially lower fluences, thanks to the higher sensitivity of HXMT.
MXB~221021 lies toward the long-duration tail of the distribution, consistent with previous findings for radio-associated events \citep{HXMT21FRB200428, GBM_catalog_2022_Oct, Cai22BurstCatalog}.
Its fluence falls within the bulk of the HXMT sample, while its relatively low flux contributes to its longer apparent duration.

Figure~\ref{fig:correlation} shows the relations between CPL spectral parameters and burst properties.
For non-radio-associated bursts, the spectral hardness (e.g., peak energy $E_{\rm p}$ for CPL fits) appears to show a weak positive trend with fluence, suggesting that more energetic bursts tend to have harder spectra, albeit with substantial scatter.
Within this framework, MXB~221021 exhibits comparatively high spectral hardness (especially in $E_{\rm c}$) relative to the non-radio-associated population, qualitatively consistent with earlier radio-associated events (e.g., FRB-MXB~200428).
Such behavior can be accommodated in models invoking enhanced resonant Compton scattering by magnetospheric electrons \citep{Younes2021magnetar, Yang2021FRB}.
However, the hardness and fluence of MXB~221021 do not follow the weak hardness--fluence trend suggested by the bulk sample, hinting that the emission conditions (e.g., geometry, plasma environment, or energy dissipation channel) in this RB-associated MXB may differ from those of typical MXBs.

\section{Summary and Conclusion} \label{sec:summary}

SGR~J1935+2154 has been one of the most active magnetars in the past decade.
In this paper, we report a joint observation by \textit{Insight}-HXMT and the KM40m telescope during the renewed activity of SGR~J1935+2154 in October 2022.
Using a refined burst-search pipeline on the LE, ME, and HE data, we identify 60 magnetar X-ray bursts (MXBs) occurring within the time intervals that overlap the KM40m radio observations.
We perform a uniform temporal and time-integrated spectral analysis for this burst sample.
For each burst, we fit several commonly used spectral models---including PL, BB, OTTB, CPL, and hybrid thermal+non-thermal combinations---and select the preferred model following the adequacy and AICc-based criteria described in Section~\ref{sec:res-and-discuss}.
A substantial fraction of bursts require a thermal component (BB and/or OTTB), indicating that thermal emission is prevalent in this activity episode.

Among the 60 MXBs, we identify MXB~221021 as the X-ray counterpart candidate of a unique radio burst detected by KM40m (RB~221021).
The time-integrated spectrum of MXB~221021 is best described by a PL model with photon index $\alpha = 1.86\pm0.09$, implying a non-thermal and relatively soft spectrum compared with the bulk of the sample.
We estimate the burst rate during the overlapping observation window and evaluate the chance-coincidence probability for an MXB to occur within the temporal vicinity of RB~221021.
The inferred probability is low, providing statistical support for a physical association between MXB~221021 and RB~221021.

A notable feature of this candidate association is the apparent delay between the radio burst and the X-ray burst.
The radio burst occurs $\sim$1~s after the X-ray peak, corresponding to $\sim 1/3$ of the spin period of SGR~J1935+2154.
While the precise value depends on the adopted timing reference (e.g., peak versus onset) and the timing uncertainty budget (including barycentering and radio dedispersion), the phase-scale offset is qualitatively different from previously reported associations from the same source, where the radio emission is nearly simultaneous with the X-ray burst or aligned with a sub-pulse within the burst envelope (e.g., MXB/FRB~200428).
This relatively large delay challenges models that predict near-simultaneity between coherent radio emission and the prompt X-ray spike, unless additional geometric and/or propagation effects are invoked.

MXB~221021 is also relatively faint in terms of flux and fluence compared to previously reported radio-associated MXBs.
Indeed, its X-ray fluence is below the typical sensitivity thresholds of operational all-sky monitors such as Fermi/GBM and GECAM, highlighting the importance of pointed, high-sensitivity observations.
In the radio band, RB~221021 is likewise much weaker than the radio bursts associated with earlier MXB events.
We estimate a radio-to-X-ray fluence ratio of $\mathcal{F}_{\rm R}/\mathcal{F}_{\rm X}\sim4\times10^{-9}$, which is orders of magnitude below that of previously reported RB-MXB associations.
This suggests that, at least in some cases, an X-ray burst can be accompanied by an extremely weak radio burst, far below the energetics of typical FRBs.
Such events help bridge the phenomenological gap between ordinary radio pulses observed in the pulsar phase and bright FRB-like bursts, supporting the notion of a continuum in magnetar radio-burst energies (e.g., \citealt{Israel2021}).

Finally, leveraging the uniform spectral fits, we examine the distributions and correlations among burst properties and spectral parameters.
Thanks to the high sensitivity of \textit{Insight}-HXMT, our sample extends to lower burst energetics than those commonly detected by all-sky monitors.
We find a weak positive association between burst fluence and duration, and a tentative trend between spectral hardness and fluence, consistent with the broad behaviors observed in earlier magnetar burst studies.
For MXB~221021, both the duration and spectral hardness lie toward the right tail of the non-radio-associated burst distributions, in line with previous findings for radio-associated events \citep{HXMT21FRB200428, Cai22BurstCatalog, GBM_catalog_2022_Oct}.

In general, MXB~221021 shares several key properties with previously identified radio-associated MXBs, but is distinguished by its comparatively low X-ray and radio energetics and, most importantly, by the apparent $\sim$second-scale radio delay.
If confirmed with additional events, such delayed associations would imply that the radio and X-ray emission are not always produced co-temporally and may originate from distinct regions or physical processes.
Future simultaneous monitoring with higher radio sensitivity and broader frequency coverage, together with high-throughput X-ray observations, will be essential to determine whether the delay is a generic feature under certain magnetospheric conditions and to place stronger constraints on the emission geometry and radiative mechanisms.

\section*{Acknowledgements}
We acknowledge the support by
the National Key R\&D Program of China (2021YFA0718500), %
the Strategic Priority Research Program of the Chinese Academy of Sciences (Grant No.
XDB0550300%
),
the National Natural Science Foundation of China (Grant Nos. 12273042, %
12494572, %
12373047, %
12333007, %
12303045%
), and the China's Space Origins Exploration Program.
This work made use of data from the \textit{Insight}-HXMT mission, funded by the CNSA and CAS.

\clearpage
\newpage
\global\pdfpageattr\expandafter{\the\pdfpageattr/Rotate 90}
\begin{longrotatetable}
\centering
\begin{deluxetable*}{cccccccccccccccc}
\label{tab:burst-analysis}
\tablecaption{X-ray Bursts detected during the simultaneous observation of \sgr{} by HXMT and KM40m. \label{bust_table}}
\tablewidth{2pt}
\tabletypesize{\scriptsize}
\tablehead{
\multicolumn{4}{c}{X-ray Burst Information} & \multicolumn{1}{c}{Fluence $^\mathrm{a}$} & \multicolumn{1}{c}{Selected} & \multicolumn{3}{c}{OTTB, PL, or CPL} & 
\multicolumn{3}{c}{BB, or BB+BB} & \multicolumn{4}{c}{OTTB/PL/CPL + BB} \\
\cmidrule(lr){1-4} \cmidrule(r){5-5} \cmidrule(lr){7-9} \cmidrule(lr){10-12} \cmidrule(lr){13-16}
ID & Burst $T_0$ & $T_\mathrm{start}$ & $\Delta T$ & $\mathcal{F}$ & Model $^\mathrm{b}$ & $\alpha$ & $E_{\rm p}$, or ${E_{\rm c}}^\mathrm{c}$ & $W$/df $^\mathrm{d}$ & $kT_1$ & $kT_2$ & $W$/df $^\mathrm{e}$ & $\alpha$ & $E_{\rm p}$, or ${E_{\rm c}}^\mathrm{c}$ & $kT$ & $W$/df $^\mathrm{f}$ \\ 
& (UTC) & (s) & (s) & \multicolumn{1}{c}{($10^{-8}$ erg cm$^{-2}$)} & & & (keV) & & (keV) & (keV) & & & (keV) & (keV) & 
}
\startdata
1 & 2022-10-15T15:40:28.625 & -0.028 & 0.121 & $1.95_{-0.13}^{+0.13}$ & OTTB+BB & $1.51_{-0.11}^{+0.10}$ & $34_{-4}^{+6}$ & 60.9/47 & $2.37_{-0.13}^{+0.15}$ & $14.3_{-0.9}^{+0.9}$ & 46.4/46 & \nodata & $54_{-7}^{+9}$ & $2.23_{-0.19}^{+0.20}$ & 40.9/46 \\
2 & 2022-10-16T10:33:30.975 & -0.036 & 0.066 & $0.45_{-0.06}^{+0.06}$ & PL & $2.14_{-0.11}^{+0.11}$ & \nodata & 53.6/44 & $1.83_{-0.35}^{+0.37}$ & $12.4_{-1.6}^{+1.8}$ & 52.1/42 & \nodata & $35_{-7}^{+10}$ & $1.24_{-0.30}^{+0.39}$ & 48.9/42 \\
3$^{g}$ & 2022-10-16T10:43:17.000 & 0.046 & 0.732 & $4.5_{-0.5}^{+0.5}$ & OTTB & \nodata & $33.1_{-3.4}^{+3.9}$ & 88.5/49 & \nodata & \nodata & \nodata & \nodata & \nodata & \nodata & \nodata \\
4 & 2022-10-16T12:03:06.000 & -0.028 & 0.056 & $0.51_{-0.07}^{+0.08}$ & OTTB+BB & $2.05_{-0.12}^{+0.12}$ & \nodata & 52.3/43 & $2.26_{-0.28}^{+0.36}$ & $14.4_{-1.7}^{+2.0}$ & 45.4/41 & \nodata & $52_{-13}^{+22}$ & $2.11_{-0.34}^{+0.41}$ & 45.0/41 \\
5 & 2022-10-16T12:17:36.545 & -0.006 & 0.012 & $0.34_{-0.05}^{+0.05}$ & PL & $2.29_{-0.15}^{+0.14}$ & \nodata & 32.3/44 & $2.0_{-0.4}^{+0.6}$ & $11.6_{-1.5}^{+1.9}$ & 32.5/42 & \nodata & $34_{-8}^{+14}$ & $1.8_{-0.5}^{+0.6}$ & 30.4/42 \\
6$^{h}$ & 2022-10-16T13:08:00.900 & -0.006 & 0.044 & $0.40_{-0.05}^{+0.05}$ & OTTB & \nodata & $50_{-12}^{+19}$ & 23.2/21 & $12.7_{-1.3}^{+1.6}$ & \nodata & 28.6/21 & $2.0_{-0.5}^{+0.6}$ & \nodata & $12_{-4}^{+5}$ & 21.8/19 \\
7 & 2022-10-16T13:26:08.400 & -0.131 & 0.162 & $0.43_{-0.06}^{+0.07}$ & PL & $1.97_{-0.15}^{+0.14}$ & \nodata & 48.2/43 & $0.97_{-0.24}^{+0.28}$ & $11.1_{-1.4}^{+1.7}$ & 42.8/41 & \nodata & $44_{-12}^{+21}$ & $0.90_{-0.28}^{+0.34}$ & 44.8/41 \\
8 & 2022-10-16T13:45:47.500 & -0.238 & 0.443 & $2.55_{-0.13}^{+0.13}$ & OTTB & \nodata & $39.7_{-3.0}^{+3.4}$ & 48.0/47 & $2.78_{-0.24}^{+0.28}$ & $13.1_{-0.7}^{+0.8}$ & 56.7/45 & \nodata & $45_{-5}^{+6}$ & $3.1_{-1.0}^{+1.3}$ & 45.8/45 \\
9 & 2022-10-16T15:02:57.200 & -0.023 & 0.096 & $0.87_{-0.07}^{+0.07}$ & CPL & $1.54_{-0.17}^{+0.16}$ & $19.8_{-3.3}^{+3.4}$ & 43.2/43 & $2.26_{-0.20}^{+0.22}$ & $11.4_{-1.0}^{+1.1}$ & 46.5/42 & \nodata & $32_{-5}^{+7}$ & $2.0_{-0.4}^{+0.4}$ & 40.5/42 \\
10 & 2022-10-17T05:42:37.175 & -0.013 & 0.154 & $1.06_{-0.07}^{+0.08}$ & OTTB & \nodata & $28.6_{-2.7}^{+3.2}$ & 49.1/45 & $2.56_{-0.30}^{+0.42}$ & $11.7_{-1.0}^{+1.1}$ & 54.5/43 & \nodata & $35_{-5}^{+6}$ & $2.3_{-0.6}^{+1.1}$ & 45.4/43 \\
11 & 2022-10-17T07:38:50.700 & -0.046 & 0.130 & $1.29_{-0.09}^{+0.09}$ & BB+BB & \nodata & $33.1_{-2.6}^{+3.0}$ & 49.6/44 & $1.92_{-0.22}^{+0.23}$ & $11.3_{-0.6}^{+0.6}$ & 36.1/42 & $2.33_{-0.17}^{+0.20}$ & \nodata & $11.0_{-0.7}^{+0.7}$ & 37.5/42 \\
12 & 2022-10-17T08:41:43.025 & -0.299 & 0.556 & $1.42_{-0.12}^{+0.14}$ & PL+BB & $0.27_{-0.30}^{+0.26}$ & $36.8_{-2.8}^{+3.2}$ & 54.8/46 & $1.7_{-0.6}^{+0.5}$ & $10.2_{-0.6}^{+0.7}$ & 49.5/45 & $2.09_{-0.27}^{+0.49}$ & \nodata & $10.0_{-0.7}^{+0.7}$ & 47.3/45 \\
13$^{h}$ & 2022-10-17T09:46:06.000 & -0.107 & 0.573 & $3.15_{-0.14}^{+0.15}$ & OTTB & \nodata & $41.8_{-3.2}^{+3.6}$ & 24.8/25 & $1.9_{-0.5}^{+0.7}$ & $12.1_{-0.5}^{+0.5}$ & 25.6/23 & $2.14_{-0.18}^{+0.25}$ & \nodata & $11.5_{-1.1}^{+1.1}$ & 18.4/23 \\
14 & 2022-10-17T10:46:03.580 & -0.014 & 0.023 & $0.29_{-0.07}^{+0.08}$ & OTTB & \nodata & $8.3_{-1.2}^{+1.5}$ & 41.8/42 & $1.3_{-0.4}^{+0.5}$ & $4.6_{-0.8}^{+1.7}$ & 41.0/40 & \nodata & \nodata & \nodata & \nodata \\
15$^{g,h,*}$ & 2022-10-17T11:14:42.000 & 0.071 & 0.792 & $61.5_{-2.2}^{+2.2}$ & CPL & $0.30_{-0.10}^{+0.10}$ & $33.3_{-1.3}^{+1.4}$ & 24.4/27 & $4.94_{-0.23}^{+0.23}$ & $12.1_{-0.4}^{+0.4}$ & 29.6/26 & \nodata & \nodata & \nodata & \nodata \\
16$^{h}$ & 2022-10-17T11:19:24.200 & -0.024 & 0.066 & $0.44_{-0.06}^{+0.06}$ & PL & $2.14_{-0.24}^{+0.25}$ & \nodata & 27.8/21 & $4.0_{-0.6}^{+0.6}$ & $29_{-7}^{+10}$ & 23.3/19 & $1.4_{-0.5}^{+0.4}$ & \nodata & $3.0_{-1.2}^{+1.0}$ & 22.7/19 \\
17$^{h}$ & 2022-10-17T11:21:27.700 & -0.205 & 0.390 & $1.66_{-0.11}^{+0.12}$ & PL & $1.92_{-0.10}^{+0.10}$ & \nodata & 15.2/23 & $4.3_{-0.6}^{+0.6}$ & $20.0_{-2.4}^{+3.3}$ & 18.4/21 & \nodata & \nodata & \nodata & \nodata \\
18$^{h}$ & 2022-10-17T11:30:16.150 & -0.266 & 0.267 & $0.68_{-0.07}^{+0.08}$ & OTTB & \nodata & $54_{-11}^{+16}$ & 21.5/22 & $12.3_{-1.1}^{+1.2}$ & \nodata & 22.0/22 & $1.0_{-2.2}^{+0.8}$ & \nodata & $10.2_{-1.4}^{+2.1}$ & 17.5/20 \\
19 & 2022-10-17T11:39:01.200 & -0.080 & 0.158 & $1.47_{-0.10}^{+0.11}$ & OTTB+BB & \nodata & $19.0_{-1.2}^{+1.3}$ & 48.8/45 & $2.48_{-0.16}^{+0.17}$ & $10.4_{-0.8}^{+0.8}$ & 51.9/43 & \nodata & $25.5_{-3.2}^{+4.0}$ & $2.53_{-0.32}^{+0.34}$ & 41.1/43 \\
20 & 2022-10-17T11:52:19.750 & -0.058 & 0.093 & $0.44_{-0.05}^{+0.06}$ & OTTB & \nodata & $39_{-8}^{+11}$ & 52.5/44 & $1.80_{-0.30}^{+0.44}$ & $13.2_{-1.5}^{+1.8}$ & 51.1/42 & \nodata & $49_{-12}^{+18}$ & $1.6_{-0.4}^{+0.5}$ & 48.9/42 \\
21 & 2022-10-17T11:52:51.500 & -0.024 & 0.432 & $1.29_{-0.09}^{+0.10}$ & OTTB & \nodata & $31.8_{-3.2}^{+3.8}$ & 53.0/47 & $2.25_{-0.24}^{+0.26}$ & $11.6_{-1.0}^{+1.1}$ & 58.3/45 & \nodata & $38_{-6}^{+8}$ & $1.7_{-0.7}^{+0.8}$ & 50.3/45 \\
22$^{h}$ & 2022-10-17T12:58:41.800 & -0.024 & 0.082 & $0.32_{-0.06}^{+0.06}$ & OTTB & \nodata & $16.7_{-3.1}^{+4.3}$ & 29.9/21 & $3.4_{-0.8}^{+0.9}$ & $9.9_{-1.9}^{+2.8}$ & 28.2/19 & \nodata & \nodata & \nodata & \nodata \\
23$^{h}$ & 2022-10-17T13:00:11.000 & -0.068 & 0.360 & $5.13_{-0.21}^{+0.22}$ & OTTB+BB & $1.75_{-0.17}^{+0.16}$ & $14_{-7}^{+5}$ & 18.7/24 & $4.55_{-0.24}^{+0.23}$ & $15.2_{-1.0}^{+1.2}$ & 14.5/23 & \nodata & $39_{-4}^{+6}$ & $4.2_{-0.4}^{+0.4}$ & 12.9/23 \\
24$^{g,*}$ & 2022-10-17T13:15:44.000 & 0.207 & 0.320 & $8.9_{-0.5}^{+0.6}$ & CPL & $0.68_{-0.07}^{+0.07}$ & $26.1_{-1.4}^{+1.5}$ & 75.0/49 & $2.64_{-0.15}^{+0.18}$ & $9.70_{-0.30}^{+0.32}$ & 85.9/48 & \nodata & \nodata & \nodata & \nodata \\
25 & 2022-10-17T13:33:53.500 & -0.022 & 0.065 & $0.30_{-0.05}^{+0.05}$ & PL & $2.04_{-0.21}^{+0.19}$ & \nodata & 34.8/44 & $1.82_{-0.32}^{+0.44}$ & $18.7_{-3.2}^{+4.2}$ & 32.1/42 & \nodata & $100_{-40}^{140}$ & $1.7_{-0.4}^{+0.5}$ & 31.7/42 \\
26 & 2022-10-17T13:51:54.630 & -0.008 & 0.137 & $0.40_{-0.06}^{+0.06}$ & PL & $1.93_{-0.17}^{+0.16}$ & \nodata & 51.7/43 & $1.94_{-0.30}^{+0.39}$ & $15.3_{-2.4}^{+3.1}$ & 46.7/41 & $1.45_{-0.30}^{+0.28}$ & \nodata & $1.85_{-0.33}^{+0.44}$ & 45.3/41 \\
27 & 2022-10-17T13:55:00.400 & -0.061 & 0.203 & $0.37_{-0.05}^{+0.06}$ & OTTB & \nodata & $13.3_{-1.8}^{+2.3}$ & 43.4/42 & $1.80_{-0.45}^{+0.34}$ & $7.7_{-1.9}^{+2.7}$ & 41.9/40 & \nodata & $20_{-6}^{+12}$ & $1.5_{-0.6}^{+0.5}$ & 40.8/40 \\
28 & 2022-10-17T13:56:25.500 & -0.070 & 0.346 & $0.91_{-0.08}^{+0.08}$ & OTTB+BB & $1.52_{-0.19}^{+0.18}$ & $22.3_{-3.5}^{+4.4}$ & 54.8/43 & $1.75_{-0.15}^{+0.16}$ & $9.9_{-1.0}^{+1.1}$ & 45.6/42 & \nodata & $35_{-6}^{+8}$ & $1.59_{-0.20}^{+0.21}$ & 44.8/42 \\
29 & 2022-10-17T14:00:08.500 & -0.147 & 0.420 & $2.51_{-0.12}^{+0.12}$ & BB+BB & \nodata & \nodata & \nodata & $2.15_{-0.20}^{+0.21}$ & $12.0_{-0.4}^{+0.5}$ & 59.3/42 & $-0.9_{-0.7}^{+0.6}$ & $45.8_{-2.5}^{+2.6}$ & $2.01_{-0.28}^{+0.26}$ & 57.3/41 \\
30$^{h}$ & 2022-10-17T14:30:36.500 & -0.463 & 0.608 & $4.79_{-0.22}^{+0.22}$ & PL+BB & \nodata & $26.8_{-1.4}^{+1.5}$ & 22.8/26 & $5.7_{-0.4}^{+0.4}$ & $15.0_{-1.8}^{+2.4}$ & 23.1/24 & $2.24_{-0.15}^{+0.14}$ & \nodata & $7.7_{-0.6}^{+0.6}$ & 15.5/24 \\
31$^{h}$ & 2022-10-17T14:34:42.600 & -0.109 & 0.170 & $0.48_{-0.07}^{+0.07}$ & PL & $1.98_{-0.22}^{+0.22}$ & \nodata & 32.9/21 & $4.6_{-0.9}^{+1.0}$ & $17.8_{-3.4}^{+5.4}$ & 30.1/19 & \nodata & $75_{-32}^{+188}$ & $4.4_{-1.4}^{+1.4}$ & 31.1/19 \\
32$^{h}$ & 2022-10-17T14:36:10.650 & -0.038 & 0.340 & $1.46_{-0.11}^{+0.12}$ & OTTB & \nodata & $65_{-10}^{+13}$ & 26.0/23 & $2.8_{-0.7}^{+0.9}$ & $15.7_{-1.1}^{+1.3}$ & 24.2/21 & $1.89_{-0.22}^{+0.50}$ & \nodata & $15.2_{-2.8}^{+2.9}$ & 22.4/21 \\
33$^{h}$ & 2022-10-17T14:39:32.975 & -0.026 & 0.040 & $0.149_{-0.026}^{+0.029}$ & BB & \nodata & $36_{-9}^{+14}$ & 27.3/21 & $11.8_{-1.7}^{+2.0}$ & \nodata & 26.2/21 & \nodata & \nodata & \nodata & \nodata \\
34 & 2022-10-17T15:04:25.000 & -0.282 & 0.683 & $2.93_{-0.17}^{+0.18}$ & PL+BB & \nodata & $37.1_{-2.3}^{+2.6}$ & 57.1/48 & $1.68_{-0.19}^{+0.20}$ & $10.5_{-0.4}^{+0.4}$ & 49.8/46 & $2.00_{-0.11}^{+0.13}$ & \nodata & $9.9_{-0.6}^{+0.6}$ & 41.3/46 \\
35 & 2022-10-17T15:11:41.200 & -0.505 & 0.705 & $1.59_{-0.16}^{+0.16}$ & PL+BB & \nodata & $37_{-4}^{+5}$ & 73.0/47 & \nodata & \nodata & \nodata & $1.75_{-0.13}^{+0.15}$ & \nodata & $7.8_{-0.8}^{+0.9}$ & 66.1/45 \\
36 & 2022-10-17T15:14:07.250 & 0.108 & 0.653 & $9.44_{-0.30}^{+0.32}$ & CPL+BB & \nodata & \nodata & \nodata & \nodata & \nodata & \nodata & $0.62_{-0.16}^{+0.14}$ & $23.1_{-1.3}^{+1.4}$ & $2.44_{-0.27}^{+0.41}$ & 68.0/52 \\
37 & 2022-10-17T15:22:56.600 & -0.022 & 0.256 & $2.06_{-0.10}^{+0.11}$ & OTTB & \nodata & $34.1_{-2.3}^{+2.6}$ & 52.4/47 & $2.47_{-0.19}^{+0.21}$ & $12.0_{-0.6}^{+0.7}$ & 61.7/45 & \nodata & $37_{-4}^{+4}$ & $2.5_{-0.9}^{+1.2}$ & 50.7/45 \\
38 & 2022-10-17T15:26:10.100 & -0.032 & 0.152 & $0.33_{-0.06}^{+0.06}$ & PL & $2.34_{-0.15}^{+0.15}$ & \nodata & 40.8/42 & $1.52_{-0.29}^{+0.31}$ & $9.8_{-1.6}^{+1.9}$ & 40.0/40 & \nodata & $23_{-6}^{+13}$ & $0.84_{-0.33}^{+0.75}$ & 40.0/40 \\
39 & 2022-10-17T15:32:20.500 & -0.125 & 0.481 & $0.81_{-0.07}^{+0.08}$ & BB+BB & \nodata & $26.0_{-2.6}^{+3.0}$ & 59.7/42 & $1.47_{-0.20}^{+0.22}$ & $9.7_{-0.7}^{+0.8}$ & 45.6/40 & $-0.9_{-1.0}^{+0.8}$ & $36.5_{-3.4}^{+3.6}$ & $1.37_{-0.22}^{+0.25}$ & 45.1/39 \\
40 & 2022-10-17T15:34:35.900 & -0.038 & 0.231 & $1.81_{-0.09}^{+0.10}$ & CPL & $1.32_{-0.12}^{+0.11}$ & $23.3_{-1.9}^{+2.1}$ & 31.4/44 & $2.16_{-0.15}^{+0.16}$ & $10.1_{-0.6}^{+0.7}$ & 54.3/43 & \nodata & $28.6_{-2.8}^{+3.3}$ & $1.9_{-0.4}^{+0.4}$ & 30.4/43 \\
41 & 2022-10-17T15:40:10.400 & -0.012 & 0.092 & $0.64_{-0.05}^{+0.06}$ & CPL & $-5.2_{-1.6}^{+1.4}$ & $58.1_{-2.9}^{+3.2}$ & 33.0/32 & $15.2_{-0.9}^{+1.0}$ & \nodata & 41.1/33 & \nodata & \nodata & \nodata & \nodata \\
42$^{h}$ & 2022-10-17T16:14:28.000 & -0.190 & 0.358 & $0.53_{-0.07}^{+0.08}$ & OTTB & \nodata & $67_{-18}^{+30}$ & 15.2/22 & $14.3_{-1.9}^{+2.1}$ & \nodata & 20.1/22 & $1.9_{-0.6}^{+1.2}$ & \nodata & $16_{-7}^{+6}$ & 15.7/20 \\
43$^{h,*}$ & 2022-10-17T16:17:17.000 & 0.014 & 0.483 & $8.80_{-0.32}^{+0.33}$ & CPL & $2.40_{-0.13}^{+0.12}$ & $91_{-24}^{+45}$ & 24.2/24 & \nodata & \nodata & \nodata & \nodata & \nodata & \nodata & \nodata \\
44$^{g,h,*}$ & 2022-10-17T16:18:19.250 & -0.175 & 0.669 & $12.2_{-0.6}^{+0.7}$ & CPL & $1.53_{-0.18}^{+0.18}$ & $16_{-5}^{+4}$ & 47.9/25 & \nodata & \nodata & \nodata & $2.53_{-0.09}^{+0.09}$ & \nodata & $7.1_{-0.6}^{+0.6}$ & 50.4/24 \\
45 & 2022-10-18T15:03:07.975 & -0.018 & 0.050 & $0.23_{-0.04}^{+0.05}$ & OTTB & \nodata & $16.2_{-3.2}^{+4.5}$ & 66.4/43 & $2.07_{-0.33}^{+0.46}$ & $10.5_{-2.0}^{+2.5}$ & 60.7/41 & \nodata & $31_{-10}^{+22}$ & $2.0_{-0.4}^{+0.5}$ & 61.1/41 \\
46$^{h}$ & 2022-10-19T06:24:50.200 & -0.018 & 0.087 & $0.78_{-0.07}^{+0.07}$ & OTTB & \nodata & $27.1_{-3.2}^{+3.9}$ & 28.9/21 & \nodata & \nodata & \nodata & \nodata & \nodata & \nodata & \nodata \\
47 & 2022-10-19T07:08:19.770 & -0.018 & 0.029 & $0.24_{-0.04}^{+0.05}$ & PL & $2.28_{-0.20}^{+0.19}$ & \nodata & 36.9/42 & $1.86_{-0.35}^{+0.44}$ & $16.4_{-2.7}^{+3.5}$ & 30.9/40 & \nodata & $72_{-27}^{+72}$ & $1.8_{-0.4}^{+0.5}$ & 32.7/40 \\
48$^{h}$ & 2022-10-19T07:55:19.700 & -0.057 & 0.115 & $0.51_{-0.06}^{+0.07}$ & PL & $1.60_{-0.22}^{+0.22}$ & \nodata & 17.6/21 & $4.2_{-1.1}^{+1.1}$ & $25_{-4}^{+6}$ & 16.4/19 & \nodata & $170_{-80}^{440}$ & $2.9_{-1.3}^{+1.7}$ & 16.2/19 \\
49 & 2022-10-19T08:35:46.300 & -0.093 & 0.426 & $1.76_{-0.13}^{+0.13}$ & PL+BB & \nodata & $31.1_{-2.3}^{+2.7}$ & 57.1/46 & $1.61_{-0.19}^{+0.19}$ & $10.0_{-0.5}^{+0.6}$ & 48.6/44 & $2.08_{-0.11}^{+0.13}$ & \nodata & $9.4_{-0.7}^{+0.7}$ & 45.5/44 \\
50 & 2022-10-19T08:41:40.150 & -0.006 & 0.015 & $0.20_{-0.04}^{+0.04}$ & PL & $2.54_{-0.22}^{+0.20}$ & \nodata & 35.8/40 & $3.2_{-0.4}^{+0.4}$ & $19_{-5}^{+8}$ & 33.0/38 & $2.27_{-0.58}^{+0.34}$ & \nodata & $3.1_{-0.8}^{+0.8}$ & 32.9/38 \\
51$^{h}$ & 2022-10-19T09:32:26.690 & -0.010 & 0.037 & $0.79_{-0.08}^{+0.09}$ & BB+BB & \nodata & \nodata & \nodata & $3.2_{-0.7}^{+0.6}$ & $16.0_{-1.4}^{+1.7}$ & 27.3/20 & \nodata & \nodata & \nodata & \nodata \\
52 & 2022-10-19T09:47:18.250 & -0.021 & 0.069 & $1.32_{-0.10}^{+0.11}$ & OTTB+BB & $1.47_{-0.14}^{+0.13}$ & $32_{-4}^{+5}$ & 44.1/44 & $2.46_{-0.20}^{+0.23}$ & $13.8_{-0.9}^{+1.0}$ & 44.1/43 & \nodata & $46_{-6}^{+8}$ & $2.31_{-0.30}^{+0.33}$ & 37.5/43 \\
53 & 2022-10-19T11:31:24.400 & -0.072 & 0.085 & $0.66_{-0.09}^{+0.10}$ & PL+BB & $1.75_{-0.15}^{+0.14}$ & \nodata & 31.8/44 & $2.20_{-0.30}^{+0.39}$ & $20.2_{-2.8}^{+3.7}$ & 27.4/42 & $1.20_{-0.27}^{+0.26}$ & \nodata & $2.10_{-0.33}^{+0.41}$ & 23.7/42 \\
54 & 2022-10-19T13:19:54.100 & -0.031 & 0.192 & $1.12_{-0.10}^{+0.10}$ & OTTB & \nodata & $18.9_{-1.7}^{+2.0}$ & 40.1/45 & $1.98_{-0.20}^{+0.20}$ & $7.9_{-0.9}^{+1.0}$ & 35.2/43 & \nodata & $24_{-4}^{+5}$ & $2.0_{-0.4}^{+0.4}$ & 37.2/43 \\
55$^{h}$ & 2022-10-20T09:26:41.250 & -0.053 & 0.076 & $0.42_{-0.06}^{+0.07}$ & OTTB & \nodata & $15.2_{-2.3}^{+2.9}$ & 30.5/21 & $4.1_{-0.8}^{+0.8}$ & $10.7_{-3.1}^{+5.6}$ & 28.8/19 & \nodata & $26_{-10}^{+38}$ & $4.2_{-1.1}^{+0.8}$ & 29.2/19 \\
56 & 2022-10-21T07:49:21.150 & -0.001 & 0.026 & $0.43_{-0.05}^{+0.06}$ & PL & $1.96_{-0.16}^{+0.15}$ & \nodata & 51.1/44 & \nodata & \nodata & \nodata & \nodata & \nodata & \nodata & \nodata \\
57$^{\star}$ & 2022-10-21T10:01:45.250 & -0.302 & 1.157 & $0.90_{-0.12}^{+0.12}$ & PL & $1.86_{-0.09}^{+0.09}$ & \nodata & 67.6/51 & $1.71_{-0.20}^{+0.21}$ & $13.0_{-2.2}^{+2.5}$ & 65.1/49 & \nodata & $66_{-20}^{+39}$ & $1.50_{-0.26}^{+0.27}$ & 62.0/49 \\
58 & 2022-10-22T09:28:00.500 & -0.020 & 0.049 & $0.32_{-0.06}^{+0.07}$ & BB+BB & \nodata & $12.1_{-2.0}^{+2.7}$ & 49.0/44 & $2.43_{-0.31}^{+0.37}$ & $12.8_{-2.6}^{+3.3}$ & 40.9/42 & \nodata & $35_{-12}^{+24}$ & $2.37_{-0.34}^{+0.40}$ & 41.9/42 \\
59 & 2022-10-22T10:00:06.750 & 0.026 & 0.239 & $7.04_{-0.26}^{+0.26}$ & OTTB+BB & \nodata & \nodata & \nodata & \nodata & \nodata & \nodata & \nodata & $30.6_{-2.0}^{+2.2}$ & $3.14_{-0.15}^{+0.15}$ & 54.2/52 \\
60 & 2022-10-28T11:45:12.775 & -0.029 & 0.071 & $0.57_{-0.08}^{+0.10}$ & BB+BB & \nodata & \nodata & \nodata & $2.4_{-0.5}^{+0.7}$ & $15.1_{-1.2}^{+1.4}$ & 46.1/40 & \nodata & $56_{-11}^{+16}$ & $2.1_{-0.5}^{+0.8}$ & 46.8/40

\enddata
\tablecomments{\\
$^\mathrm{a}$ 8--200$\,$keV energy fluence of the selected model. \\
$^\mathrm{b}$ Model with the minimum of AICc. \\
$^\mathrm{c}$ For CPL with $\alpha \geq 2$, we report $E_\mathrm{c}$ instead of $E_\mathrm{p}$. Note that for OTTB model, $E_\mathrm{p}$ and $E_\mathrm{c}$ are the same as $kT_\mathrm{e}$. \\
$^\mathrm{d}$ $W$-statistics and the degrees of freedom for the OTTB, PL, or CPL model. \\
$^\mathrm{e}$ $W$-statistics and the degrees of freedom for the BB, or BB + BB model. \\
$^\mathrm{f}$ $W$-statistics and the degrees of freedom for the OTTB/PL/CPL + BB model. \\
$^{\mathrm{g}}$ Bright bursts caused HE to suffer from saturation, leading to inaccurate spectral results. An additional factor was fitted to the HE data of these bursts to reduce inaccuracy. \\
$^{\mathrm{h}}$ LE data is unavailable due to the saturation caused by the solar X-ray flux reflected by Earth. \\
$^*$ Bursts also detected by \textit{Fermi}/GBM.\\
$^\star$ MXB221021 associated with RB 221021 detected by KM40m.
}
\end{deluxetable*}
\end{longrotatetable}

\global\pdfpageattr\expandafter{\the\pdfpageattr/Rotate 0}
\clearpage
\newpage

\bibliography{refs}
\bibliographystyle{aasjournalv7}

\end{document}